\documentclass{juliacon}
\usepackage{amsmath,booktabs,mathtools,amssymb}
\usepackage{tikz}
\usetikzlibrary{arrows.meta}
\usepackage{pgfplots}
\usepackage{pgfplotstable}
\usepackage{balance}
\pgfplotsset{compat=1.18}

\usepackage{siunitx}
\usepackage{microtype}
\makeatletter
\renewcommand{\fnum@table}{Table \thetable}
\makeatother

\definecolor{TolMutedBlue}{HTML}{332288}
\definecolor{TolMutedCyan}{HTML}{88CCEE}
\definecolor{TolMutedTeal}{HTML}{44AA99}
\definecolor{TolMutedGreen}{HTML}{117733}
\definecolor{TolMutedOlive}{HTML}{999933}
\definecolor{TolMutedSand}{HTML}{DDCC77}
\definecolor{TolMutedRose}{HTML}{CC6677}
\definecolor{TolMutedWine}{HTML}{882255}
\definecolor{TolMutedPurple}{HTML}{AA4499}
\hypersetup{%
    colorlinks=true,%
    urlcolor=TolMutedRose,%
    linkcolor=TolMutedBlue,%
    citecolor=TolMutedGreen%
}

\newcommand{\scorio}{\texttt{Scorio.jl}}

\newcolumntype{L}[1]{>{\raggedright\arraybackslash}p{#1}}

\begin{document}


\title{Scorio.jl: A Julia package for ranking stochastic responses}

\author[1]{Mohsen Hariri}
\author[2]{Michael Hinczewski}
\author[1]{Vipin Chaudhary}
\affil[1]{Department of Computer and Data Sciences, Case Western Reserve University, Cleveland, OH, USA}
\affil[2]{Department of Physics, Case Western Reserve University, Cleveland, OH, USA}

\keywords{Julia, Ranking, Stochastic evaluation, Numerical analysis, Scientific computing, Large Language Models}

\hypersetup{
pdftitle = {Scorio.jl: A Julia package for ranking stochastic responses},
pdfsubject = {JuliaCon 2025 Proceedings},
pdfauthor = {Mohsen Hariri, Michael Hinczewski, Vipin Chaudhary},
pdfkeywords = {Julia, Ranking, Stochastic evaluation, Numerical analysis, Scientific computing, Large Language Models},
}

\maketitle

\begin{abstract}
    \href{https://github.com/mohsenhariri/scorio}{\scorio{}} is a Julia package for evaluating and ranking systems from repeated responses to shared tasks. It provides a common tensor-based interface for direct score-based, pairwise, psychometric, voting, graph, and listwise methods, so the same benchmark can
    be analyzed under multiple ranking assumptions. We describe the package design, position it relative to existing Julia tools, and report pilot experiments on synthetic rank recovery, stability under limited trials, and runtime scaling.
\end{abstract}

\section{Summary}

When a benchmark involves stochastic outputs---language models sampled multiple times per prompt \cite{snell2025testtime,hariri2026ranking}, randomized software tests, or repeated simulation runs---the natural data object is a three-dimensional tensor \(R \in \{0,\ldots,C\}^{L\times M\times N}\) recording \(L\) systems on \(M\) tasks over \(N\) trials. Different ranking methods consume different views of this tensor: averages, pairwise win counts, per-task orderings, or latent trait estimates. Moving between these views typically means moving between separate tools.

\scorio{} keeps the tensor as the single input and handles the view conversions internally. For binary-response benchmarks, a user can pass the same \texttt{R} to \texttt{avg}, \texttt{bradley\_terry}, \texttt{rasch}, \texttt{borda}, or \texttt{pagerank} and get back comparable rank vectors without touching the surrounding pipeline; weighted score-based methods such as \texttt{bayes} also support graded outcomes. The package therefore serves two roles: day-to-day leaderboard construction and methodological studies of how rankings shift when the ranking rule changes. Julia's multiple dispatch and dense-array primitives make it practical to keep these heterogeneous methods behind one API \cite{bezanson2017julia}.

\section{Statement of need}

Ranking under repeated stochastic observation is common across several fields. In educational testing, examinees answer the same items multiple times or under adaptive schedules \cite{BockAitkin1981,LairdWare1982}. In software benchmarking, nondeterminism from hardware scheduling or random initialization means that a single run is rarely conclusive. In language-model evaluation, metrics like Pass@k aggregate multiple samples per task, and limited sample budgets can produce unstable rankings \cite{chen2021evaluating,hariri2025dontpassk}.

In all these cases, the analyst faces questions that go beyond ``which system scored highest on average'': How many trials are needed before the ranking stabilizes? Do pairwise skill models and direct score averages agree? Does the top system change when ties are handled differently? Can prior information from earlier runs be folded in?

Answering these questions today requires stitching together separate tools, each with its own input format. \scorio{} removes that friction. It treats the repeated response tensor as the canonical data structure and lets the user swap ranking methods with a single function call. This is useful whenever the ranking rule itself is part of the analysis rather than a fixed post-processing step.

\section{State of the field}

Several Julia packages handle parts of this problem. StatsBase provides general-purpose ranking and tied-ranking utilities but does not operate on repeated benchmark tensors \cite{statsbasejl}. RankAggregation.jl aggregates multiple score columns into a consensus ordering \cite{rankaggregationjl}. Ranking.jl focuses on binary comparison records and includes draft Elo, Bradley--Terry, and TrueSkill implementations \cite{rankingjl}. TrueSkillThroughTime.jl specializes in temporally evolving skill estimation \cite{trueskillthroughtimejl}. These are useful packages, but each addresses a narrower slice of the problem.

\scorio{} differs in two ways. First, its primitive input is a repeated response tensor over systems, tasks, and trials under stochastic sampling. Second, it brings together evaluation-based, pairwise, probabilistic, psychometric, voting, graph, and listwise rankers behind one interface. That combination is the package's main software contribution: it turns ranking-method comparison from a custom integration problem into an ordinary part of benchmark analysis.

\section{Software design}

\subsection{Response tensors and a shared API}

For evaluating a single system, \scorio{} uses outcome matrices \(E \in \{0,\ldots,C\}^{M\times N}\), where \(M\) is the number of tasks, \(N\) the number of trials, and \(C\) the maximum category label (\(C+1\) outcome categories in total). For ranking multiple systems, it uses response tensors \(R \in \{0,\ldots,C\}^{L\times M\times N}\), where \(L\) is the number of systems. Two-dimensional inputs are promoted automatically to a single-trial tensor, so one-shot benchmarks use the same code path. When outcomes are graded, score-based methods can consume them directly via rubric weights.

Different method families consume different views of \(R\). Pointwise methods work with trial means \(\bar r_{\ell m} = \frac{1}{N}\sum_{n=1}^{N} R_{\ell mn}\) or rubric-weighted scores. Pairwise methods work with induced decisive-win counts \(W_{\ell,\ell'} = \lvert\{(m,n) : R_{\ell mn} > R_{\ell' mn}\}\rvert\), with ties tracked separately when needed. Voting and listwise methods use the per-task ordering induced by \((R_{1mn},\ldots,R_{Lmn})\). Graph methods build a weighted comparison graph. IRT methods estimate latent abilities and task difficulties from binary outcomes. All these conversions happen inside the package; the user passes tensors in the same \((L,M,N)\) format.

\subsection{Score construction, tie handling, and priors}

Score construction is separated from rank assignment. Most methods return continuous scores, and a shared tie-aware ranking layer converts those scores to ranks. This avoids duplicating tie logic and makes the rank semantics (competition, dense, ordinal, fractional) explicit.

The package also exposes explicit priors for MAP-style methods,
including Gaussian, Laplace, Cauchy, uniform, custom, and empirical priors. This is useful when previous benchmark runs or baseline
evaluations should inform the current ranking, and it encourages
users to state their assumptions rather than burying them in custom
code.

\subsection{Illustrative Julia workflow}

A typical workflow starts from an integer tensor \texttt{R} with shape \texttt{(L, M, N)}. The same object can be passed to several rankers without any reshaping. Listing~\ref{lst:workflow} shows the shared API.

\begin{lstlisting}[language=julia,basicstyle=\ttfamily\scriptsize,breaklines=true,breakatwhitespace=false,columns=fullflexible,keepspaces=true,frame=single,showstringspaces=false,caption={Illustrative \scorio{} workflow using the shared tensor interface.},label={lst:workflow}]
using Scorio

# R has size (L, M, N) for systems, tasks, and trials
r_bayes, s_bayes = bayes(R; return_scores=true)
r_bt,    s_bt    = bradley_terry(R; return_scores=true)
r_rasch, theta   = rasch_mml(R; return_scores=true)
r_pr,    s_pr    = pagerank(R; return_scores=true)

# A graded evaluation for one system across tasks and trials
E = [0 1 2 2 1;
     1 1 0 2 2]
w = [0.0, 0.5, 1.0]
mu, sigma = bayes(E, w)

# Convert a score vector into several tie-aware rank views
rank_views = rank_scores(s_bayes)
\end{lstlisting}

\subsection{Method families}

Table~\ref{tab:families} lists the ranking families currently available in \scorio{} together with the data view each family uses. The evaluation-based family includes \texttt{avg}, \texttt{bayes}, and several Pass@k variants \cite{chen2021evaluating,hariri2025dontpassk}. Sequential rating models (Elo, Glicko, TrueSkill) process induced pairwise matches \cite{elo1978,glickman1999,herbrich2006trueskill}. Paired-comparison models such as Bradley--Terry, Davidson, and Rao--Kupper estimate strength parameters from aggregated win/tie counts \cite{BradleyTerry1952,davidson1970bties,rao1967ties}, alongside Thompson sampling and a Metropolis MCMC ranker \cite{Thompson1933,Metropolis1953}. Item-response theory models (Rasch, 2PL, 3PL, MML, and a dynamic variant) estimate latent ability jointly with task difficulty \cite{rasch1960rasch,birnbaum1968latent,Mislevy1986,BockAitkin1981,verhelst1993dynamicrasch}. Voting rules cover Borda, Copeland, Schulze, ranked pairs, Kemeny--Young, Nanson, Baldwin, and majority judgment \cite{Borda1781,Copeland1951,Condorcet1785,Schulze2010,Tideman1987,Kemeny1959,Young1977,Nanson1883,Baldwin1926,BalinskiLaraki2011}. Graph and spectral methods include PageRank, Rank Centrality, AlphaRank, a Nash meta-game solver, SerialRank, and HodgeRank \cite{page1999pagerank,vigna2016spectral,negahban2017rankcentrality,omidshafiei2019alpharank,balduzzi2018reevaluating,fogel2016serialrank,jiang2009hodgerank}. Finally, listwise choice models include Plackett--Luce, Davidson--Luce, and Bradley--Terry--Luce \cite{plackett1975permutations,firth2019davidsonluce,luce1959choice}.

\begin{table*}[t]
\centering
\caption{Ranking families in \scorio{} and the data view each uses. The table shows representative methods; several families expose additional keyword variants.}
\label{tab:families}
\small
\begin{tabular}{L{0.18\textwidth}L{0.45\textwidth}L{0.27\textwidth}}
\toprule
Family & Representative methods & Primary data view \\
\midrule
Evaluation-based & \texttt{avg}, \texttt{bayes}, \texttt{pass\_at\_k}, \texttt{pass\_hat\_k}, \texttt{g\_pass\_at\_k\_tau}, \texttt{mg\_pass\_at\_k} & Per-system repeated outcomes and rubric-weighted scores \\
Pointwise & \texttt{inverse\_difficulty} & Per-task solve rates with task reweighting \\
Sequential rating & \texttt{elo}, \texttt{glicko}, \texttt{trueskill} & Stream of induced pairwise matches over tasks and trials \\
Paired-comparison & \texttt{bradley\_terry}, \texttt{bradley\_terry\_davidson}, \texttt{rao\_kupper}, \texttt{thompson}, \texttt{bayesian\_mcmc} & Aggregated win and tie counts between systems \\
Voting & \texttt{borda}, \texttt{copeland}, \texttt{schulze}, \texttt{ranked\_pairs}, \texttt{kemeny\_young}, \texttt{nanson}, \texttt{baldwin}, \texttt{majority\_judgment} & Per-task orderings induced by repeated responses \\
IRT & \texttt{rasch}, \texttt{rasch\_2pl}, \texttt{rasch\_3pl}, \texttt{rasch\_mml}, \texttt{dynamic\_irt} & Latent system ability and task difficulty \\
Graph and spectral & \texttt{pagerank}, \texttt{rank\_centrality}, \texttt{alpharank}, \texttt{nash}, \texttt{serial\_rank}, \texttt{hodge\_rank} & Weighted comparison graph over systems \\
Listwise and choice & \texttt{plackett\_luce}, \texttt{davidson\_luce}, \texttt{bradley\_terry\_luce} & Winner and loser sets for each task and trial \\
\bottomrule
\end{tabular}
\end{table*}

This breadth is the main software contribution: users can compare ranking assumptions on the same data without moving between incompatible tools.

\section{Experimental evaluation}

\begin{figure*}[t]
\centering
\includegraphics[width=0.49\textwidth]{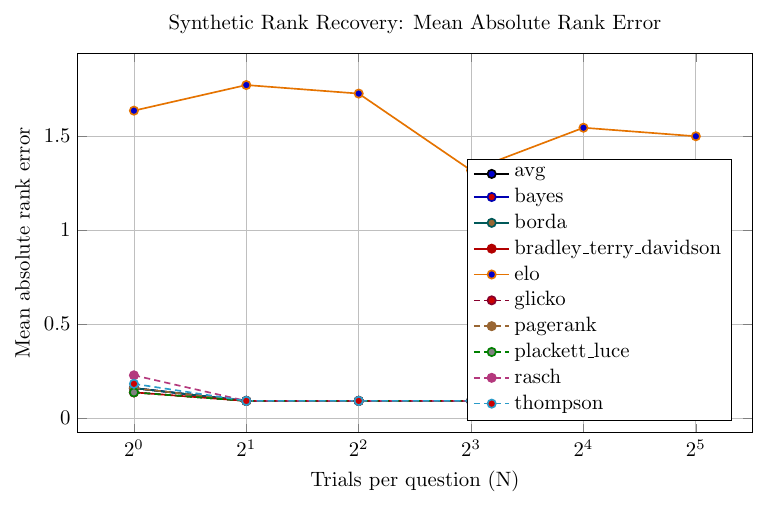}\hfill
\includegraphics[width=0.49\textwidth]{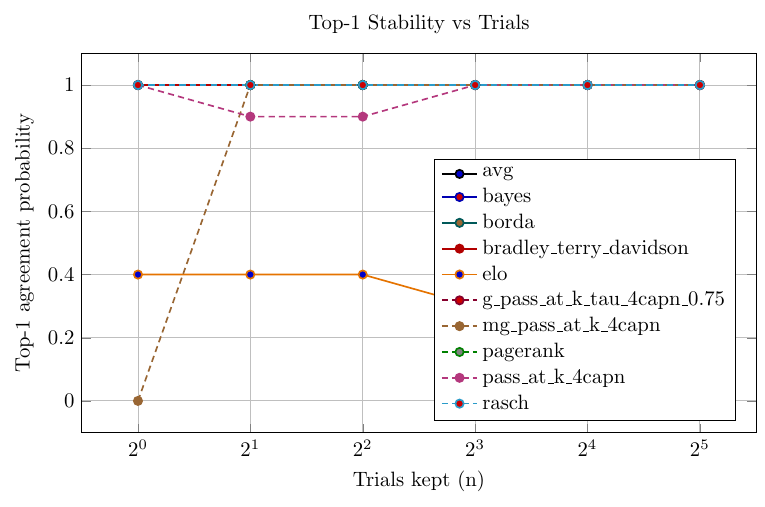}
\caption{Complementary views of the same experiments. \textbf{Left:} mean absolute rank error versus trials. \textbf{Right:} top-1 agreement under limited trials. mG-Pass@k degenerates to all ties at \(n=1\); Elo remains unreliable throughout.}
\label{fig:additional-pilot}
\end{figure*}

We test three questions: (i)~whether representative methods recover a known synthetic ranking, (ii)~how stable the rankings are when only a few trials are available, and (iii)~what the practical runtime costs look like. The results below come from the \texttt{pilot} profile. Figure~\ref{fig:additional-pilot} provides complementary views via mean absolute rank error and top-1 stability.

\subsection{Synthetic rank recovery}

We generate binary response tensors from a Rasch-style latent ability model with \(L=11\) systems, \(M=500\) tasks, \(N \in \{1,2,4,8,16,32\}\) trials, and four random seeds. The ground-truth ability vector contains one intentional tie (two systems share the same ability). Figure~\ref{fig:pilot-recovery-stability} (left) and Table~\ref{tab:recovery} summarize the results.

Most methods recover the latent order well even at \(N=1\): Kendall \(\tau_b\) ranges from \(0.963\) (Rasch) to \(0.982\) (BT-Davidson, Plackett--Luce). By \(N=32\), every non-Elo method reaches \(\tau_b=0.991\) with a mean absolute rank error of~\(0.091\).

Elo is the clear outlier (\(\tau_b = 0.624\) at \(N=1\), \(0.661\) at \(N=32\)). This is expected: Elo updates ratings sequentially using task--trial pairs as matches, and the ordering of those matches introduces path-dependent noise that does not wash out even with more trials.

One negative result is worth noting: none of the tested methods recovered the intentional ground-truth tie, suggesting that exact tie recovery is harder than recovering the coarse rank order in this setting. To address that, the package includes a tie-aware ranking layer that can be applied to any score vector, so users can experiment with different tie thresholds.

\begin{table}[t]
\centering
\caption{Synthetic rank recovery (mean over 4 seeds). Top-1 recovery is \(1.0\) for every method except Elo. No method recovered the ground-truth tie.}
\label{tab:recovery}
\footnotesize
\begin{tabular}{lcccc}
\toprule
 & \multicolumn{2}{c}{$N=1$} & \multicolumn{2}{c}{$N=32$} \\
\cmidrule(lr){2-3}\cmidrule(lr){4-5}
Method & $\tau_b$ & MAE & $\tau_b$ & MAE \\
\midrule
avg & 0.979 & 0.159 & 0.991 & 0.091 \\
bayes & 0.979 & 0.159 & 0.991 & 0.091 \\
BT-Davidson & 0.982 & 0.136 & 0.991 & 0.091 \\
Rasch & 0.963 & 0.227 & 0.991 & 0.091 \\
PageRank & 0.979 & 0.159 & 0.991 & 0.091 \\
Plackett--Luce & 0.982 & 0.136 & 0.991 & 0.091 \\
Elo & 0.624 & 1.636 & 0.661 & 1.500 \\
\bottomrule
\end{tabular}
\end{table}

\subsection{Stability under limited trials}

\begin{figure*}[t]
\centering
\includegraphics[width=0.49\textwidth]{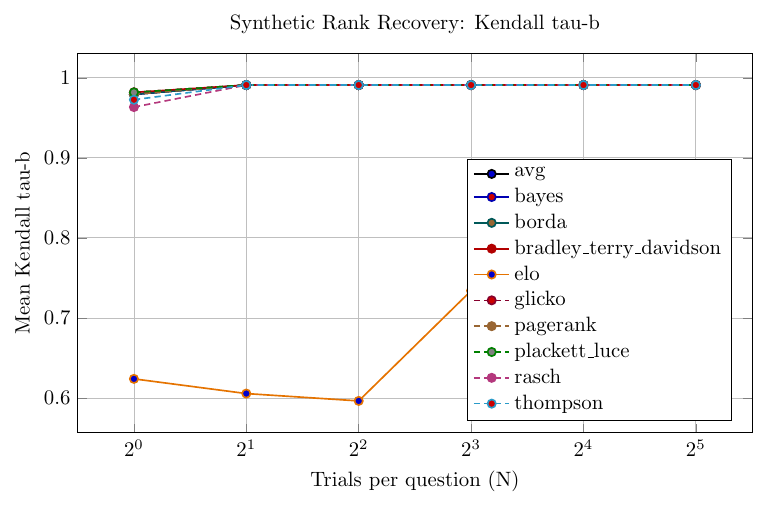}\hfill
\includegraphics[width=0.49\textwidth]{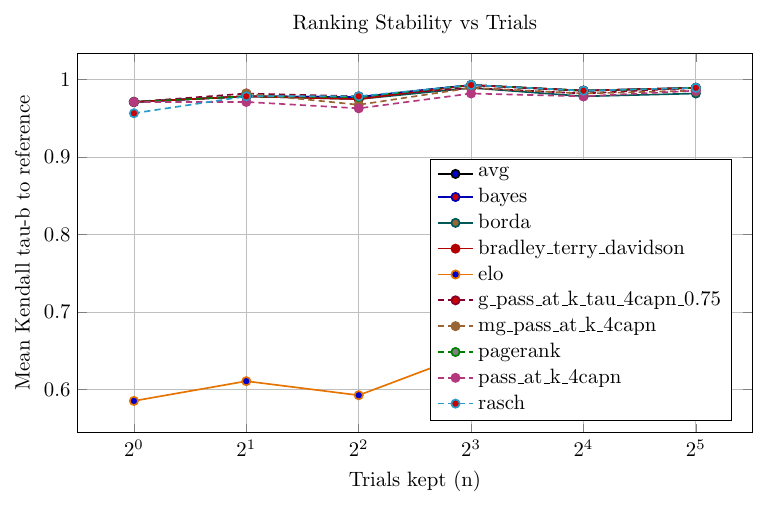}
\caption{\textbf{Left:} Kendall \(\tau_b\) between predicted and ground-truth rankings as the number of trials grows. Most methods cluster near~\(1.0\) quickly; Elo does not. \textbf{Right:} stability of limited-trial rankings relative to a \texttt{bayes} reference at \(N_{\max}=64\). Method differences are most visible in the small-\(n\) regime.}
\label{fig:pilot-recovery-stability}
\end{figure*}

For ten synthetic datasets with \(N_{\max}=64\) trials each, we take the \texttt{bayes} ranking at \(N_{\max}\) as a reference and recompute rankings from the first \(n \in \{1,2,4,8,16,32\}\) trials only. Figure~\ref{fig:pilot-recovery-stability} (right) and Table~\ref{tab:stability} show the results.

Most methods are already close to the reference at \(n=1\). The direct score-based methods (\texttt{avg}, \texttt{bayes}, g-Pass@k), PageRank, and BT-Davidson all start at \(\tau_b \approx 0.971\) with perfect top-1 agreement, and they reach about \(0.989\) by \(n=32\). Rasch starts lower (\(0.956\)) but catches up. Standard Pass@k is slightly less stable in the smallest-\(n\) regime but reaches full top-1 agreement for \(n \ge 8\).

The edge case is mG-Pass@k, which at \(n=1\) collapses to all ties (undefined \(\tau_b\), zero top-1 agreement). By \(n=2\) it joins the other score-based methods. Elo stays near \(\tau_b \approx 0.59\) throughout and never reaches reliable top-1 agreement. These results show that method choice matters most when trials are scarce---exactly the regime where practitioners have the least data to guide that choice.

\begin{table}[t]
\centering
\caption{Ranking stability relative to a \texttt{bayes} reference at \(N_{\max}=64\) (mean over 10 seeds). For mG-Pass@k at \(n=1\), all systems are tied, so \(\tau_b\) is undefined.}
\label{tab:stability}
\footnotesize
\begin{tabular}{lcccc}
\toprule
 & \multicolumn{2}{c}{$n=1$} & \multicolumn{2}{c}{$n=32$} \\
\cmidrule(lr){2-3}\cmidrule(lr){4-5}
Method & $\tau_b$ & Top-1 & $\tau_b$ & Top-1 \\
\midrule
avg & 0.971 & 1.000 & 0.989 & 1.000 \\
bayes & 0.971 & 1.000 & 0.989 & 1.000 \\
Pass@k & 0.971 & 1.000 & 0.985 & 1.000 \\
g-Pass@k & 0.971 & 1.000 & 0.989 & 1.000 \\
mG-Pass@k & -- & 0.000 & 0.985 & 1.000 \\
Rasch & 0.956 & 1.000 & 0.989 & 1.000 \\
PageRank & 0.971 & 1.000 & 0.989 & 1.000 \\
Elo & 0.585 & 0.400 & 0.589 & 0.200 \\
\bottomrule
\end{tabular}
\end{table}

\subsection{Runtime scaling}

\begin{figure*}[t]
\centering
\includegraphics[width=\textwidth]{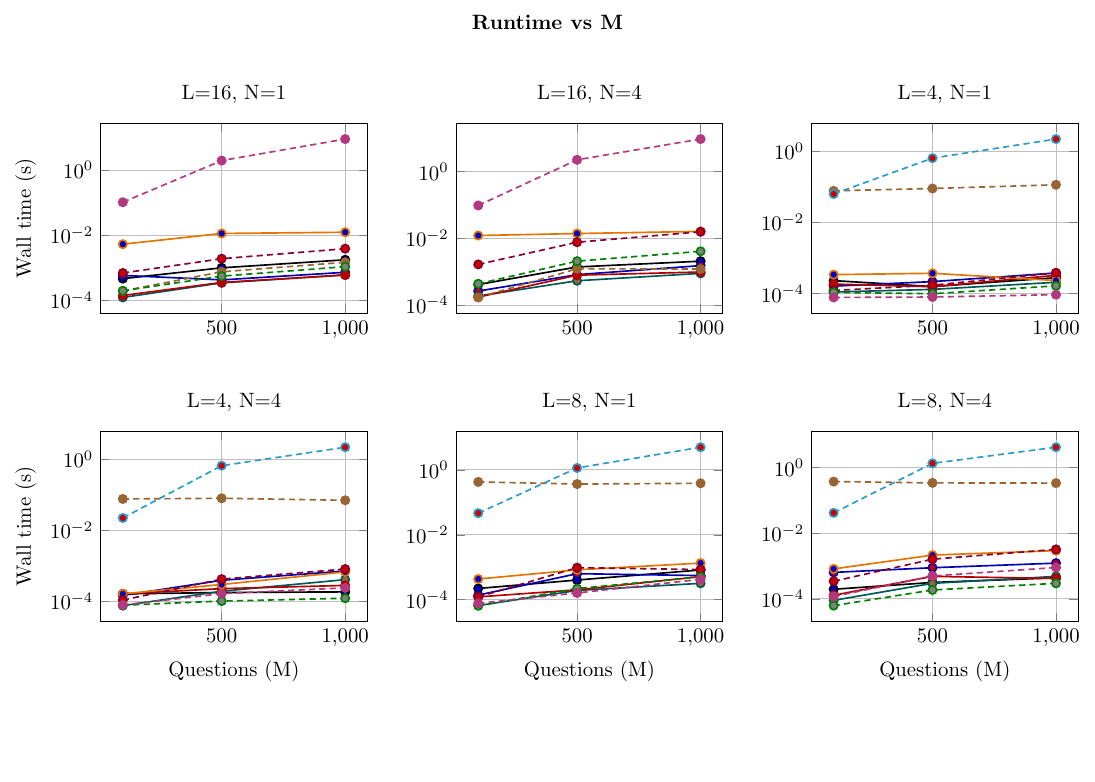}
\caption{Runtime versus the number of tasks \(M\) for several \((L,N)\) configurations. Score-based and graph methods stay fast across the grid. Rasch and Kemeny--Young are substantially more expensive.}
\label{fig:runtime}
\end{figure*}

We benchmark wall-clock time across \(L\in\{4,8,16\}\), \(M\in\{100,500,1000\}\), and \(N\in\{1,4\}\), with one warm-up call and two timed replicates per configuration. Figure~\ref{fig:runtime} shows runtime versus~\(M\). Kemeny--Young is included only for \(L \le 8\) due to its factorial complexity.

The cost differences are large. \texttt{bayes}, Borda, PageRank, and \texttt{avg} stay below \(2 \times 10^{-3}\) s per call on every tested configuration; AlphaRank is near that threshold (max \(2.05 \times 10^{-3}\) s), and Plackett--Luce is near \(4 \times 10^{-3}\) s (max \(4.06 \times 10^{-3}\) s). Elo and Bradley--Terry are in the low-millisecond range but scale noticeably with \(L\) and \(M\). Kemeny--Young is orders of magnitude slower (\(0.34\text{--}0.42\) s at \(L=8\)), reflecting the NP-hardness of the underlying optimization. Rasch is the most expensive method tested, reaching \(8.95\) s at \((L,M,N)=(16,1000,1)\) with roughly \(12.8\) GB of allocations, because the current MML implementation uses dense quadrature.

\balance
\section{Research impact statement}

\scorio{} lowers the cost of studying how benchmark conclusions depend on the ranking rule. In stochastic software benchmarks, repeated simulation studies, educational assessment, and any setting with shared tasks and repeated observations, the ranking method is a modeling choice that can change the outcome. By providing a common interface for multiple ranking families, the package makes it straightforward to run sensitivity analyses and compare assumptions---turning what would otherwise be a custom integration effort into a single function call.

\section{Conclusion}

\scorio{} provides a common Julia interface for ranking repeated responses to shared tasks. Its central contribution is software infrastructure: direct score-based, pairwise, psychometric, voting, graph, and listwise methods can be applied to the same tensor-valued input without rebuilding the surrounding workflow. In the pilot experiments reported here, most representative methods recover and stabilize the tested synthetic rankings quickly, whereas the runtime measurements reveal substantial differences in computational cost across method families. These results position \scorio{} as a practical workbench for comparative ranking analysis under repeated stochastic evaluation.

\balance

\bibliographystyle{juliacon}
\bibliography{ref.bib}

\end{document}